\documentclass[12pt]{article}
\usepackage{graphicx}
\def\Journal#1#2#3#4{{#1} {\bf #2}, #3 (#4)}

\def\PLB{Phys. Lett. B}

\def\PRD{Phys. Rev. D}
\def\ZPC{Z. Phys. C}
\def\EPJ{Eur. Phys. J. C}

\newcommand{\ra}{\rightarrow}
\newcommand{\Lam}{\Lambda_{c}^{+}}
\newcommand{\Lambar}{\bar{\Lambda}_{c}^{+}}
\newcommand{\Lamvec}{\vec{\Lambda}_{c}^{+}}
\newcommand{\Lambarvec}{\vec{\bar{\Lambda}}_{c}^{{}_{{}_+}}}

\begin{document}
\thispagestyle{empty}
\begin{flushright}
KOBE--FHD--02--02\\
January~~~2002
\end{flushright}

\vspace{1cm}
\begin{center}
\baselineskip=1cm
{\Large \bf Polarized $s$--quark Distribution \\
in Charmed Hadron Leptoproduction\footnote[2]{Talk presented at SPIN2001 
(The 3rd Circum--Pan--Pacific Symposium on ``High Energy Spin Physics''), 
Beijing, China, Octobar 2001.}}

\baselineskip=0.6cm
\vspace{1.5cm}
{\large Kazutaka ~SUDOH} \\
\vspace{0.5cm}
{\em Graduate School of Science and Technology, Kobe University, \\
Nada, Kobe 657--8501, JAPAN} \\
\vspace{0.3cm}
E-mail: {\tt sudou@radix.h.kobe-u.ac.jp}
\end{center}

\vspace{2cm}
\noindent
\begin{center}
{\bf ABSTRACT}
\end{center}

In order to extract the polarized strange quark density in proton, 
we studied the semi--inclusive $\Lam$/$\Lambar$ leptoproduction 
in charged current DIS at THERA energies. 
We indicate that measurements of the spin correlation between 
the incident proton and the produced $\Lam$/$\Lambar$ baryon gives us 
information about the polarized strange quark distribution. 

\vspace{0.5cm}
\clearpage

\section{Introduction}

Proton spin puzzle is currently one of the most challenging topics 
in high energy spin physics. 
As is well known, proton spin is composed of the spin and orbital 
angular momentum of quarks and gluons. 
The polarized parton distribution function (PDF) plays an important 
role on deep understandings of the spin structure of proton. 
However, knowledge about the polarized sea quark distribution remains 
still poor. 
In order to understand the spin structure of proton, we need more 
information about the polarized sea quark distribution functions. 
Recently, HERMES group has reported \cite{HERMES99} that direct 
measurement of the strange sea is required to explain the violation of 
the Ellis--Jaffe sum rule \cite{Ellis74}. 
So far, there are several parametrization models of the strange quark 
distribution. 
Though the most simplest case is to assume the flavor ${\rm SU(3)}_f$, 
a new parametrization taken account of the violation of ${\rm SU(3)}_f$ 
is also recently proposed \cite{Leader99}. 

The study of heavy quark production in deep inelastic scattering (DIS) 
is one of the promising way to access the parton density in proton, 
since heavy quark is not main constituents of proton and is only 
produced in hard processes. 
Heavy quark leptoproduction in neutral current DIS has already measured 
by some collaborations at HERA. 
However, no measurement of heavy quark production in charged current 
DIS has been so far done at HERA due to the small cross section. 
We can expect that the cross section for charged current heavy quark 
production will be several times larger at THERA than at 
HERA \cite{Gladilin01}. 
THERA is a future high energy $ep$ collider which uses electron or 
positron from the linear collider TESLA at energies of 250 GeV to 800 
GeV and proton from HERA at energy 920 GeV. 
Thus, charged current heavy quark production is also interesting and 
challenging topics. 

In this work, to examine the polarized $s$/$\bar{s}$ quark distribution, 
we have studied semi--inclusive $\Lam$/$\Lambar$ leptoproduction 
in charged current DIS; 
$l^+ + \vec{p} \ra \bar{\nu} + \Lamvec + X$, 
$l^- + \vec{p} \ra \nu + \Lambarvec + X$, 
where arrows attached to particles mean that these particles are polarized. 
These processes might be observed in the forthcoming THERA experiments. 
Since in the naive quark model, the $\Lam$ baryon is composed of a heavy 
$c$--quark and anti--symmetrically combined light $u$ and $d$--quarks, 
we can assume the polarization of $\Lam$ baryon to be the one of charm quark. 
In addition, $\Lam$ is dominantly produced by the fragmentation of 
charm quark which is originated from strange quark through the $t$--channel 
$W^{\pm}$ boson exchange at the leading order. 
Therefore, there can be a correlation between the strange quark 
polarization and the produced charm quark polarization. 
Hence we can expect that the measurement of the spin correlation 
between the incident proton and the produced $\Lam$/$\Lambar$ gives us 
information about the polarized $s$/$\bar{s}$ quark distribution in proton. 

\section{Double Spin Asymmetry}

For above processes, we calculated the double spin asymmetry $A_{LL}^H$ 
for final state hadron specified by $H$ ($\Lam$ or $\Lambar$), 
which is given by 
\begin{equation}
A_{LL}^{H}\equiv 
\frac{[ d\sigma_{++} - d\sigma_{+-}+ d\sigma_{--} - d\sigma_{-+}]/dp_{T}}
{[ d\sigma_{++} + d\sigma_{+-}+ d\sigma_{--} + d\sigma_{-+}]/dp_{T}}
=\frac{d\Delta\sigma/dp_{T}}{d\sigma/dp_{T}} ,
\label{ALL}
\end{equation}
where $d\sigma_{+-}$, for instance, denotes that the helicity of proton 
is positive and the one of $H$ is negative. 
$p_T$ is a transverse momentum of final hadron $H$. 

For $\Lam$ production, the spin--dependent differential 
cross section at leading order level can be generally written by 
\begin{equation}
d\Delta\sigma(l^{+} p\ra \bar{\nu} \Lam X)
=\left\{{\rm U}_{cs}^2 \Delta s(x)+{\rm U}_{cd}^2 \Delta d(x)\right\}dx
\left(\frac{d\Delta\hat{\sigma}}{d\hat{t}}
\right)d\hat{t}\Delta D_c^{\Lam} (z)dz ,
\end{equation}
where $\Delta s(x)$ and $\Delta d(x)$ are the polarized 
$s$--quark and $d$--quark distribution functions, respectively.
${\rm U}_{cs}$ and ${\rm U}_{cd}$ are CKM parameters. 
$\Delta D_c^{\Lam}(z)$ is the polarized fragmentation function of 
outgoing charm quark decaying into $\Lam$. 
We used the model of Peterson {\em et al.} \cite{Peterson83} as the 
unpolarized fragmentation function \cite{PDG98}. 
Unfortunately, we have at present no data about the polarized 
fragmentation function because of lack of experimental data. 
By analogy with the study on $\Lambda$ polarization \cite{deFlorian98}, 
we took the 
following ansatz: 
\begin{equation}
\Delta D_c^{\Lam}(z)=C_c^{\Lam}(z) D_c^{\Lam}(z) , 
\end{equation}
where $C_c^{\Lam}(z)$ is the scale--independent spin transfer 
coefficient. 
Here we apply the analysis on $\Lambda$ production to $\Lam$ production 
and choose the following two models; $C_c^{\Lam}(z)=1$ (the naive 
nonrelativistic quark model) and $C_c^{\Lam}(z)=z$ (the jet 
fragmentation models \cite{Bartl80}) to evaluate cross sections. 

The double spin asymmetry $A_{LL}^{\Lam}$ is described in terms of 
the spin transfer coefficient $C_c^{\Lam}(z)$ and the ratio of 
the PDF's as 
\begin{equation}
A_{LL}^{\Lam}\propto C_c^{\Lam}(z)\frac{\left\{{\rm U}_{cs}^2 \Delta s(x)
+{\rm U}_{cd}^2 \Delta d(x)\right\}}{\left\{{\rm U}_{cs}^2 s(x)
+{\rm U}_{cd}^2 d(x)\right\}}. 
\end{equation}
Thus $A_{LL}^{\Lam}$ is proportional to linear combination of 
the polarized $s$--quark and $d$--quark distribution function. 
Since the contribution from the valence $d$--quark is large, the 
$s$--quark distribution is not so clearly extracted in $\Lam$ production 
case. 

On the contrary, $A_{LL}^{\Lambar}$ in $\Lambar$ production can be
represented as 
\begin{equation}
A_{LL}^{\Lambar}\propto C_c^{\Lam}(z)\frac{\left\{{\rm U}_{cs}^2 \Delta 
\bar{s}(x)+{\rm U}_{cd}^2 \Delta \bar{d}(x)\right\}}
{\left\{{\rm U}_{cs}^2 \bar{s}(x)+{\rm U}_{cd}^2 \bar{d}(x)\right\}} 
=C_c^{\Lam}(z)\frac{\Delta \bar{s}(x)}{\bar{s}(x)}
\simeq C_c^{\Lam}(z)\frac{\Delta s(x)}{s(x)}, 
\end{equation}
in which the $\bar{d}$--quark contribution and CKM parameters can be 
eliminated, and then $A_{LL}^{\Lambar}$ is directly proportional to the 
strange quark distribution function alone. 
Therefore, we can clearly extract the polarized strange quark 
distribution $\Delta s(x)$. 
Note that above equation is derived in both cases; the flavor 
${\rm SU(3)}_f$ case ($\Delta \bar{u}(x)=\Delta \bar{d}(x)=\Delta \bar{s}(x)$)
and the non--${\rm SU(3)}_f$ case 
($\Delta \bar{u}(x)=\Delta \bar{d}(x)=\lambda \Delta \bar{s}(x)$\footnote
{$\lambda$ represent a degree of ${\rm SU(3)}_f$ violation and is a 
parameter which should be determined from experiments \cite{Leader99}.}). 

\section{Numerical Results}

Setting a charm quark mass $m_c = 1.5$ GeV and the relevant collider 
energy $\sqrt{s}=300$ GeV, we numerically calculated the 
spin--independent and dependent differential cross sections and the 
double spin asymmetry. 
As for the PDF's, we used the GRV98 \cite{GRV98} parametrization as 
the unpolarized PDF, and the AAC \cite{AAC} and ``standard scenario'' of 
GRSV01 \cite{GRSV01} parametrizations for the polarized one. 

\begin{figure}[htbn]
\begin{center}
  \includegraphics[width=17pc, keepaspectratio]{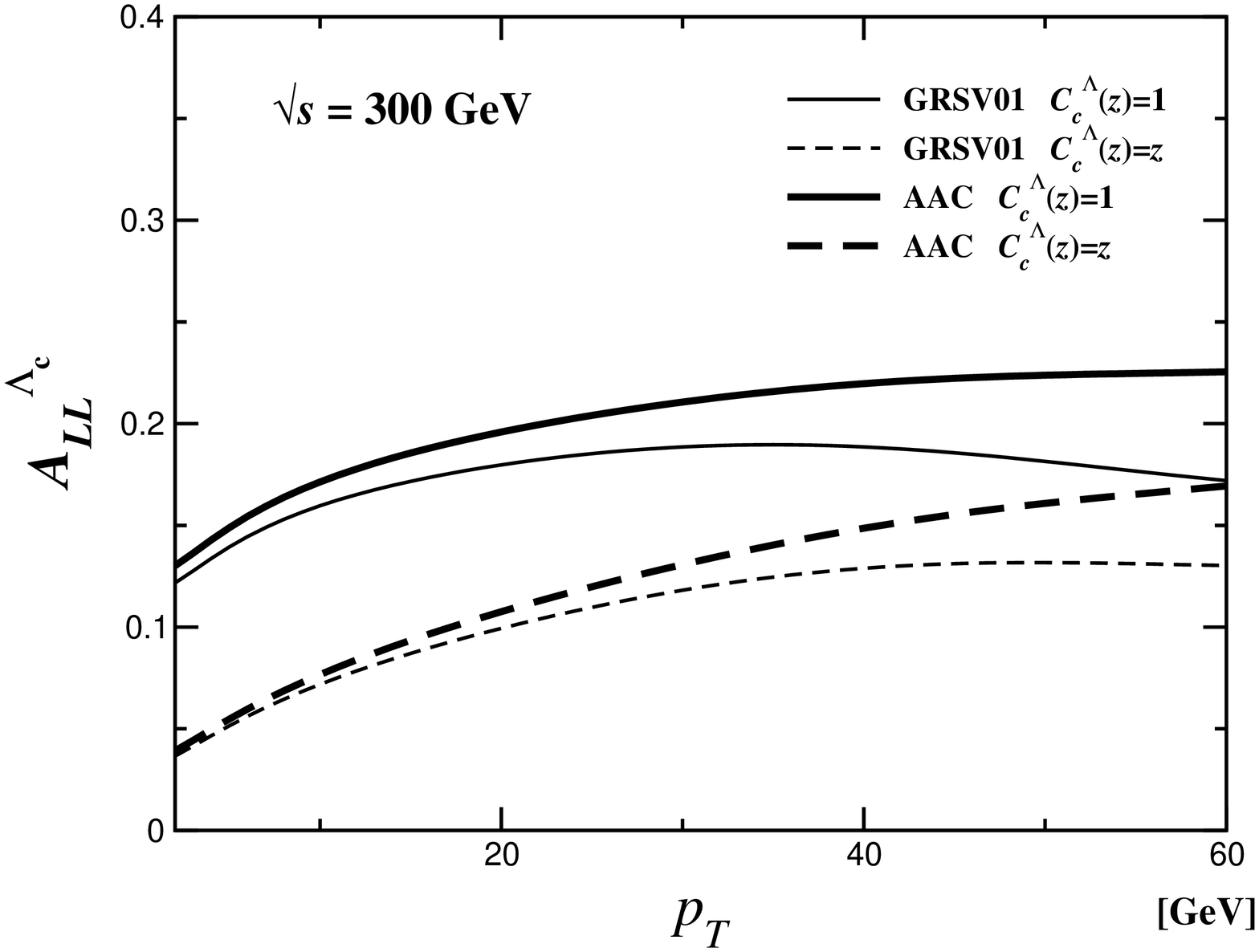}
  \includegraphics[width=17pc, keepaspectratio]{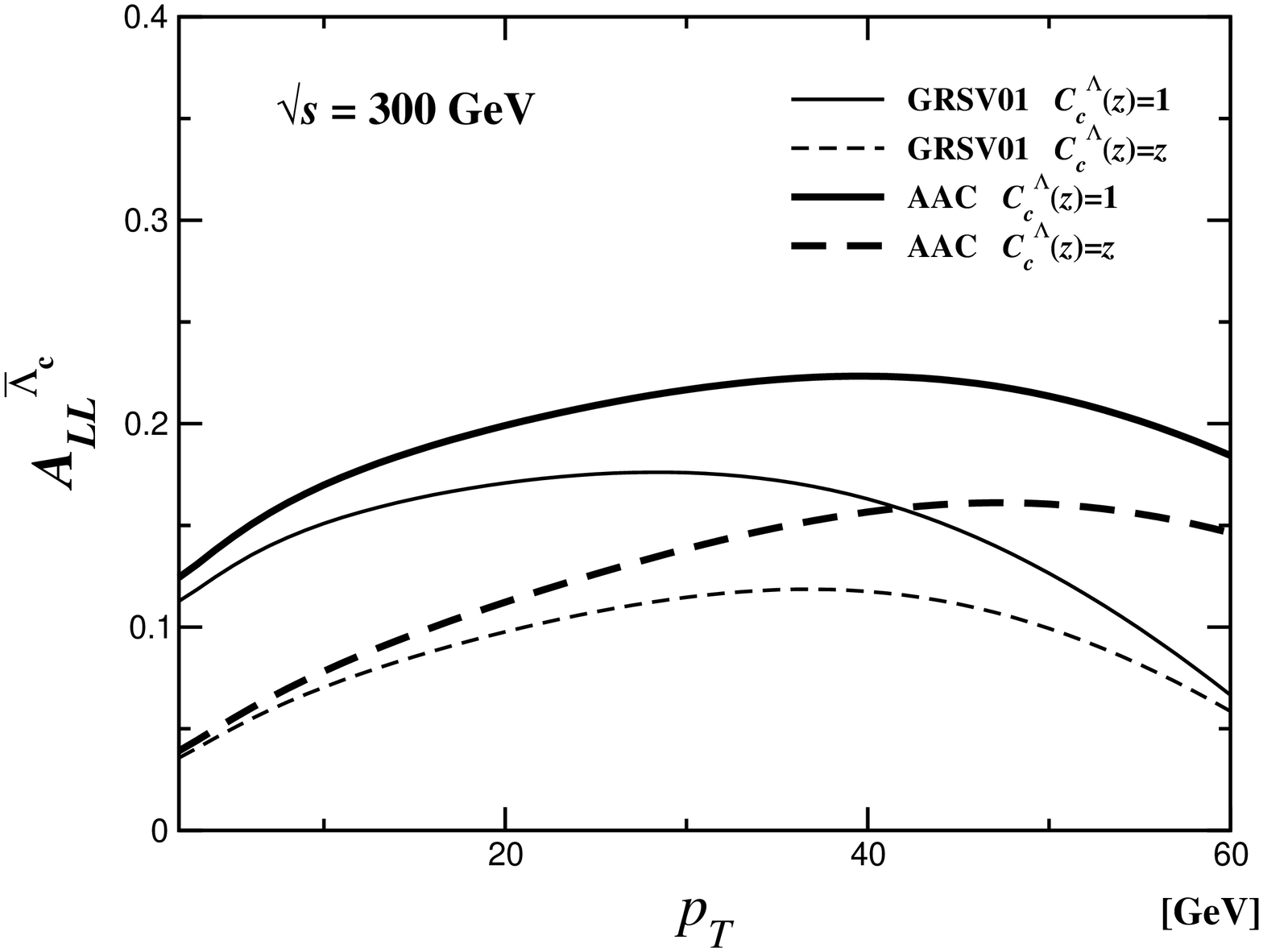}
\end{center}
\caption{
$p_T$ distribution of double spin asymmetries at $\sqrt{s}=300$ GeV
for $\Lam$ production (left panel) and 
for $\Lambar$ production (right panel). 
}
\label{fig1}
\end{figure}

We show the double spin asymmetry in Fig. \ref{fig1} as a function of 
$p_T$ at $\sqrt{s}=300$ GeV for $\Lam$ production (left panel) and for 
$\Lambar$ production (right panel). 
In order to suppress the contributions from the diffractive process and 
higher twist corrections, we have imposed the kinematical cut on $p_T$ 
as $p_T > 2$ GeV in numerical calculations. 
In figures, the bold and normal lines show the case of AAC 
parametrization and ``standard scenario'' of GRSV01 parametrization, 
respectively. 
The solid lines represent the spin transfer coefficient 
$C_c^{\Lam}(z)=1$ case, while the dashed lines represent 
$C_c^{\Lam}(z)=z$ case. 

As shown in figures, $A_{LL}^{\Lam}$ in smaller $p_T$ regions does 
not depend on the model of polarized PDF's, and is strongly affected 
by the shape of the spin transfer coefficient $C_c^{\Lam}(z)$. 
Therefore, The $\Lam$ production is effective for extracting information 
about the polarized fragmentation function $\Delta D_c^{\Lam}(z)$. 
In addition, we can directly measure the polarized strange quark 
distribution function, though the parametrization model dependence 
is not so large. 
On the other hand, for $\Lambar$ production $A_{LL}^{\Lambar}$ for 
GRSV01 parametrization rapidly decreases in larger $p_T$ regions, and 
we see big difference between two parametrization models. 
Therefore, measuring $A_{LL}^{\Lambar}$ in larger $p_T$ regions 
is quite effective for testing the model of polarized PDF's, since 
the ambiguity of the polarized fragmentation function becomes small. 

\section{Summary}

In summary, to extract information about the polarized strange quark 
distribution in proton, the semi--inclusive $\Lam$/$\Lambar$ production 
in charged current DIS in unpolarized lepton--polarized proton 
collisions was studied. 
$\Lambar$ production is most promising not only for testing the 
parametrization models of PDF's but also for directly extracting the 
polarized strange quark distribution $\Delta s(x)$ by measuring the 
double spin asymmetry $A_{LL}^{\Lambar}$. 

\vspace{0.5cm}

\end{document}